\input harvmac
\input graphicx
\input color

\def\Title#1#2{\rightline{#1}\ifx\answ\bigans\nopagenumbers\pageno0\vskip1in
\else\pageno1\vskip.8in\fi \centerline{\titlefont #2}\vskip .5in}

%
%
\ifx\includegraphics\UnDeFiNeD\message{(NO graphicx.tex, FIGURES WILL BE IGNORED)}
\def\figin#1{\vskip2in}
\else\message{(FIGURES WILL BE INCLUDED)}\def\figin#1{#1}
\fi
\def\Fig#1{Fig.~\the\figno\xdef#1{Fig.~\the\figno}\global\advance\figno
 by1}
%
%
%
%

\font\ticp=cmcsc10

\def \purge#1 {\textcolor{magenta}{#1}}
\def \new#1 {\textcolor{blue}{#1}}
\def\comment#1{}

\def\\{\cr}
\def\text#1{{\rm #1}}
\def\frac#1#2{{#1\over#2}}

\def\subsubsec#1{\noindent{\undertext {#1}}}
\def\undertext#1{$\underline{\smash{\hbox{#1}}}$}
\def\hf{{1\over 2}} 
\def\calo{{\cal O}}

\def\calb{{\cal B}}
\def\cald{{\cal D}}

\def\calh{{\cal H}}
\def\cala{{\cal A}}

\def\eg{{\it e.g.}}
\def\roughly#1{\mathrel{\raise.3ex\hbox{$#1$\kern-.75em\lower1ex\hbox{$\sim$}}}}
\font\bbbi=msbm10 
\def\mathbb#1{\hbox{\bbbi #1}}

\def\mthsu{\mathsurround=0pt  }
\def\leftrightarrowfill{$\mthsu \mathord\leftarrow\mkern-6mu\cleaders
  \hbox{$\mkern-2mu \mathord- \mkern-2mu$}\hfill
  \mkern-6mu\mathord\rightarrow$}
\def\overleftrightarrow#1{\vbox{\ialign{##\crcr\leftrightarrowfill\crcr\noalign{\kern-1pt\nointerlineskip}$\hfil\displaystyle{#1}\hfil$\crcr}}}
\overfullrule=0pt

%
%
\lref\Haag{R. Haag, {\sl Local quantum physics, fields, particles, algebras,} Springer (Berlin, 1996).}
\lref\YngIII{
  J.~Yngvason,
  ``The Role of type III factors in quantum field theory,''
Rept.\ Math.\ Phys.\  {\bf 55}, 135 (2005).
[math-ph/0411058].
}
\lref\DoGi{W. Donnelly and S.B.~Giddings, in preparation.}
\lref\ZanardiZZ{
  P.~Zanardi, D.~A.~Lidar and S.~Lloyd,
  ``Quantum tensor product structures are observable induced,''
Phys.\ Rev.\ Lett.\  {\bf 92}, 060402 (2004).
[quant-ph/0308043].
}
\lref\CavesZZ{
  C.~M.~Caves and B.~L.~Schumaker,
  ``New formalism for two-photon quantum optics. 1. Quadrature phases and squeezed states,''
Phys.\ Rev.\ A {\bf 31}, 3068 (1985).
}
\lref\BuchholzGR{
  D.~Buchholz and R.~Verch,
  ``Scaling algebras and renormalization group in algebraic quantum field theory,''
Rev.\ Math.\ Phys.\  {\bf 7}, 1195 (1995).
[hep-th/9501063].
}
\lref\Donnelly{
  W.~Donnelly,
  ``Entanglement entropy and nonabelian gauge symmetry,''
Class.\ Quant.\ Grav.\  {\bf 31}, no. 21, 214003 (2014).
[arXiv:1406.7304 [hep-th]].
}
\lref\Trieste{
  S.~B.~Giddings,
  ``Quantum mechanics of black holes,''
[hep-th/9412138].
}
\lref\locbd{
  S.~B.~Giddings and M.~Lippert,
  ``Precursors, black holes, and a locality bound,''
Phys.\ Rev.\ D {\bf 65}, 024006 (2002).
[hep-th/0103231]
\semi
  S.~B.~Giddings and M.~Lippert,
  ``The Information paradox and the locality bound,''
Phys.\ Rev.\ D {\bf 69}, 124019 (2004).
[hep-th/0402073].
}
\lref\LQGST{
  S.~B.~Giddings,
  ``Locality in quantum gravity and string theory,''
Phys.\ Rev.\ D {\bf 74}, 106006 (2006).
[hep-th/0604072].
}
\lref\Thie{
  K.~Giesel and T.~Thiemann,
  ``Algebraic Quantum Gravity (AQG). I. Conceptual Setup,''
Class.\ Quant.\ Grav.\  {\bf 24}, 2465 (2007).
[gr-qc/0607099].
}
\lref\BHQIUE{
  S.~B.~Giddings,
  ``Black holes, quantum information, and unitary evolution,''
  Phys.\ Rev.\ D {\bf 85}, 124063 (2012).
[arXiv:1201.1037 [hep-th]].
}
\lref\SGmodels{
  S.~B.~Giddings,
   ``Models for unitary black hole disintegration,''  Phys.\ Rev.\ D {\bf 85}, 044038 (2012)
[arXiv:1108.2015 [hep-th]].
}
\lref\Mathurrev{
  S.~D.~Mathur,
  ``Fuzzballs and the information paradox: A Summary and conjectures,''
[arXiv:0810.4525 [hep-th]].
}
\lref\GiShI{
  S.~B.~Giddings and Y.~Shi,
  ``Quantum information transfer and models for black hole mechanics,''
Phys.\ Rev.\ D {\bf 87}, no. 6, 064031 (2013).
[arXiv:1205.4732 [hep-th]].
}
\lref\Susstrans{
  L.~Susskind,
  ``The Transfer of Entanglement: The Case for Firewalls,''
[arXiv:1210.2098 [hep-th]].
}
\lref\UQM{
  S.~B.~Giddings,
  ``Universal quantum mechanics,''
Phys.\ Rev.\ D {\bf 78}, 084004 (2008).
[arXiv:0711.0757 [quant-ph]].
}
\lref\Hart{
  J.~B.~Hartle,
  ``Space-time quantum mechanics and the quantum mechanics of space-time,''
[gr-qc/9304006].
}
\lref\GMH{
  S.~B.~Giddings, D.~Marolf and J.~B.~Hartle,
  ``Observables in effective gravity,''
Phys.\ Rev.\ D {\bf 74}, 064018 (2006).
[hep-th/0512200].
}
\lref\Dirac{
  P.~A.~M.~Dirac,
  ``Gauge invariant formulation of quantum electrodynamics,''
Can.\ J.\ Phys.\  {\bf 33}, 650 (1955).
}
\lref\Shab{
  S.~V.~Shabanov,
  ``The Proper field of charges and gauge invariant variables in electrodynamics,''
Submitted to: J. Phys. A.
}
\lref\Heem{
  I.~Heemskerk,
  ``Construction of Bulk Fields with Gauge Redundancy,''
JHEP {\bf 1209}, 106 (2012).
[arXiv:1201.3666 [hep-th]].
}
\lref\KLdecode{
  D.~Kabat and G.~Lifschytz,
  ``Decoding the hologram: Scalar fields interacting with gravity,''
Phys.\ Rev.\ D {\bf 89}, no. 6, 066010 (2014).
[arXiv:1311.3020 [hep-th]].
}
\lref\GiMa{
  S.~B.~Giddings and D.~Marolf,
  ``A Global picture of quantum de Sitter space,''
Phys.\ Rev.\ D {\bf 76}, 064023 (2007).
[arXiv:0705.1178 [hep-th]].
}
\lref\GiSl{
  S.~B.~Giddings and M.~S.~Sloth,
  ``Fluctuating geometries, q-observables, and infrared growth in inflationary spacetimes,''
Phys.\ Rev.\ D {\bf 86}, 083538 (2012).
[arXiv:1109.1000 [hep-th]].
}
\lref\GiSr{
  S.~B.~Giddings and M.~Srednicki,
  ``High-energy gravitational scattering and black hole resonances,''
Phys.\ Rev.\ D {\bf 77}, 085025 (2008).
[arXiv:0711.5012 [hep-th]].
}
\lref\GiPo{
  S.~B.~Giddings and R.~A.~Porto,
  ``The Gravitational S-matrix,''
Phys.\ Rev.\ D {\bf 81}, 025002 (2010).
[arXiv:0908.0004 [hep-th]].
}
\lref\Erice{
  S.~B.~Giddings,
  ``The gravitational S-matrix: Erice lectures,''
[arXiv:1105.2036 [hep-th]].
}
\lref\Hawkinc{
  S.~W.~Hawking,
 ``Breakdown of Predictability in Gravitational Collapse,''
Phys.\ Rev.\ D {\bf 14}, 2460 (1976).
}
\lref\NVNLFT{
  S.~B.~Giddings,
  ``Nonviolent information transfer from black holes: A field theory parametrization,''
Phys.\ Rev.\ D {\bf 88}, no. 2, 024018 (2013).
[arXiv:1302.2613 [hep-th]].
}
\lref\GiShii{
  S.~B.~Giddings and Y.~Shi,
  ``Effective field theory models for nonviolent information transfer from black holes,''
Phys.\ Rev.\ D {\bf 89}, no. 12, 124032 (2014).
[arXiv:1310.5700 [hep-th]].
}
\lref\Gimod{
  S.~B.~Giddings,
  ``Modulated Hawking radiation and a nonviolent channel for information release,''
Phys.\ Lett.\ B {\bf 738}, 92 (2014).
[arXiv:1401.5804 [hep-th]].
}
\lref\Isit{
  S.~B.~Giddings,
  ``Is string theory a theory of quantum gravity?,''
Found.\ Phys.\  {\bf 43}, 115 (2013).
[arXiv:1105.6359 [hep-th]].
}
\lref\GaGi{
  M.~Gary and S.~B.~Giddings,
  ``Constraints on a fine-grained AdS/CFT correspondence,''
[arXiv:1106.3553 [hep-th]].
}
\lref\Maldeternal{
  J.~M.~Maldacena,
  ``Eternal black holes in anti-de Sitter,''
JHEP {\bf 0304}, 021 (2003).
[hep-th/0106112].
}
\lref\vanR{
  M.~Van Raamsdonk,
  ``Building up spacetime with quantum entanglement,''
Gen.\ Rel.\ Grav.\  {\bf 42}, 2323 (2010), [Int.\ J.\ Mod.\ Phys.\ D {\bf 19}, 2429 (2010)].
[arXiv:1005.3035 [hep-th]].
}
\lref\BLR{
  R.~Bousso, S.~Leichenauer and V.~Rosenhaus,
  ``Light-sheets and AdS/CFT,''
Phys.\ Rev.\ D {\bf 86}, 046009 (2012).
[arXiv:1203.6619 [hep-th]].
}
\lref\CzechBH{
  B.~Czech, J.~L.~Karczmarek, F.~Nogueira and M.~Van Raamsdonk,
  ``The Gravity Dual of a Density Matrix,''
Class.\ Quant.\ Grav.\  {\bf 29}, 155009 (2012).
[arXiv:1204.1330 [hep-th]].
}
\lref\BoussoMH{
  R.~Bousso, B.~Freivogel, S.~Leichenauer, V.~Rosenhaus and C.~Zukowski,
  ``Null Geodesics, Local CFT Operators and AdS/CFT for Subregions,''
Phys.\ Rev.\ D {\bf 88}, 064057 (2013).
[arXiv:1209.4641 [hep-th]].
}
\lref\HubenyWA{
  V.~E.~Hubeny and M.~Rangamani,
  ``Causal Holographic Information,''
JHEP {\bf 1206}, 114 (2012).
[arXiv:1204.1698 [hep-th]].
}
\lref\MaSu{
  J.~Maldacena and L.~Susskind,
  ``Cool horizons for entangled black holes,''
Fortsch.\ Phys.\  {\bf 61}, 781 (2013).
[arXiv:1306.0533 [hep-th]].
}
\lref\AMPS{
  A.~Almheiri, D.~Marolf, J.~Polchinski and J.~Sully,
  ``Black Holes: Complementarity or Firewalls?,''
  JHEP {\bf 1302}, 062 (2013).
  [arXiv:1207.3123 [hep-th]].
}
\lref\BHMR{
  S.~B.~Giddings,
  ``Black holes and massive remnants,''
Phys.\ Rev.\ D {\bf 46}, 1347 (1992).
[hep-th/9203059].
}
\lref\NVNL{
  S.~B.~Giddings,
  ``Nonviolent nonlocality,''
  Phys.\ Rev.\ D {\bf 88},  064023 (2013).
[arXiv:1211.7070 [hep-th]].
}
\lref\GiSh{
  S.~B.~Giddings and Y.~Shi,
  ``Quantum information transfer and models for black hole mechanics,''
Phys.\ Rev.\ D {\bf 87}, 064031 (2013).
[arXiv:1205.4732 [hep-th]].
}
\lref\AMPSS{
  A.~Almheiri, D.~Marolf, J.~Polchinski, D.~Stanford and J.~Sully,
  ``An Apologia for Firewalls,''
JHEP {\bf 1309}, 018 (2013).
[arXiv:1304.6483 [hep-th]].
}
\lref\CHR{
  H.~Casini, M.~Huerta and J.~A.~Rosabal,
  ``Remarks on entanglement entropy for gauge fields,''
Phys.\ Rev.\ D {\bf 89}, no. 8, 085012 (2014).
[arXiv:1312.1183 [hep-th]].
}
\lref\BHIUN{
  S.~B.~Giddings,
  ``Black hole information, unitarity, and nonlocality,''
Phys.\ Rev.\ D {\bf 74}, 106005 (2006).
[hep-th/0605196].
}
\lref\NLvC{
  S.~B.~Giddings,
 ``Nonlocality versus complementarity: A Conservative approach to the information problem,''
Class.\ Quant.\ Grav.\  {\bf 28}, 025002 (2011).
[arXiv:0911.3395 [hep-th]].
}
\lref\AlmheiriLWA{
  A.~Almheiri, X.~Dong and D.~Harlow,
  ``Bulk Locality and Quantum Error Correction in AdS/CFT,''
JHEP {\bf 1504}, 163 (2015).
[arXiv:1411.7041 [hep-th]].
}
\lref\RequardtKJA{
  M.~Requardt,
 ``The Thermal Aspects of Relativistic Quantum Field Theory as an Observational Window in a Deeper Layer of Quantum Space-Time or: Dirac's Revenge,''
[arXiv:1309.1351 [hep-th]]\semi
``The Incomplete Semiclassical Analysis of the Black Hole Information Paradox and its Completion via Entanglement of Radiation and Quantum Gravity Degrees of Freedom,''
[arXiv:1503.07312 [gr-qc]].
}
\Title{
\vbox{\baselineskip12pt  
}}
{\vbox{
\centerline{Hilbert space structure 
in quantum gravity:}\centerline{an algebraic perspective}
}}

\centerline{{\ticp 
Steven B. Giddings\footnote{$^\ast$}{Email address: giddings@physics.ucsb.edu}
} }
\centerline{\sl Department of Physics}
\centerline{\sl University of California}
\centerline{\sl Santa Barbara, CA 93106}
\vskip.10in
\centerline{\bf Abstract}

If quantum gravity respects the principles of quantum mechanics, suitably generalized, it may be that a more viable approach to the theory is through identifying the relevant quantum structures rather than by quantizing classical spacetime.  This viewpoint is supported by difficulties of such quantization, and by the apparent lack of a fundamental role for locality.  In finite or discrete quantum systems, important structure is provided by tensor factorizations of the Hilbert space.  However, even in local quantum field theory properties of the generic type III von Neumann algebras and of long range gauge fields indicate that factorization of the Hilbert space is problematic.  Instead it is better to focus  on the structure of the algebra of observables, and in particular on its subalgebras corresponding to regions.  This paper suggests that study of analogous algebraic structure in gravity gives an important perspective on the nature of the quantum theory. Significant departures from the subalgebra structure of local quantum field theory are found, working in the correspondence limit of long-distances/low-energies.  Particularly, there are obstacles to identifying commuting algebras of localized operators.   In addition to suggesting important properties of the algebraic structure, this and related observations pose challenges to proposals of a fundamental role for entanglement.

\vskip.3in
\Date{}

\newsec{Introduction}

Physics currently faces a foundational crisis, encountered when we attempt  to understand gravitational phenomena:
the principles of quantum mechanics, the principles of relativity, and the principle of locality are in basic conflict.  This is particularly sharply revealed by study of black hole evolution, which has no consistent description respecting these basic principles.

With these general principles  in conflict, a central question becomes what to trust.  Quantum mechanics has a theoretically rigid structure, forbidding many possible pathologies, and experience has shown it is hard to modify without introducing these pathologies.  This, together with its many experimental tests, suggests that it may be a cornerstone of physics, at least when sufficiently generalized to accommodate the possible lack of a fundamental role for space and time.

Lorentz invariance is also well-tested and hard to modify, suggesting a possible fundamental role for it when describing states corresponding to flat or nearly-flat spacetime.  

On the other hand, locality is difficult to even carefully formulate in a quantum theory of gravity. This is associated with the lack of a precise role for classical spacetime, and in turn indicates that the local invariance (general covariance) is also likely subject to modification, since diffeomorphisms assume local manifold structure.

A possible primacy of quantum mechanics over spacetime suggests that spacetime should emerge, in an approximation, from structures that are intrinsically quantum-mechanical.  If this is the case, this also suggests that one should take a different approach than the program of starting with a classical description of spacetime, via general relativity, and attempting to quantize it.  Plausibly quantum mechanics plays a fundamental role in physics, and spacetime, locality, and the equivalence principle emerge as approximations to more basic quantum structures.

What kind of additional structure can a quantum-mechanical theory have, beyond the existence of a Hilbert space?  For finite quantum systems relevant to physics, or those that are ``locally finite," with finite density of degrees of freedom, such as on a lattice, one typically has in addition a tensor factor structure, which gives a quantum notion of ``locality."  However, in the case of local quantum field theory (LQFT), as we will review, such tensor factor structures cannot be precisely defined.  Instead, a more fundamental role is played by the algebraic structure of the operators acting on the Hilbert space\Haag.  In particular, in LQFT, subalgebras of the observables may be associated to regions of spacetime, and the resulting subalgebra structure mirrors the topology of the spacetime manifold.  

Such a ``quantum mechanics first" approach, together with the need for a theory of quantum gravity to be approximated by LQFT, in the low-energy/long-distance limit, suggests that we should likewise focus on algebraic structure in seeking a fundamental theory of gravity.  This, combined with important aspects of the long-distance behavior of quantum gravity, and in particular properties of black holes and cosmology, and constraints from the S-matrix, likely provides important constraints on and clues toward the basic theory.

This paper will focus on an initial investigation of the idea of approaching quantum gravity from such an algebraic perspective, and on a possible role for such a ``quantum emergent geometry."  The next section will give a brief overview of the Hilbert space structure and algebraic structure of LQFT, in particular discussing the problems with tensor factorizations.  

Section three then turns to algebraic aspects of quantum gravity, introducing essential postulates and beginning an investigation of the structure of the relevant operator algebra.   Properties of this algebra can be inferred based on an assumption of ``correspondence," namely that the fundamental theory matches onto an effective theory description given by general relativity in the long-distance/low-energy limit.  Even this amount of information reveals novel algebraic structure, departing from the local algebraic structures of LQFT.  Specifically, gravitationally-dressed, diffeomorphism-invariant operators fail to commute at spacelike separations.  This significantly differs from the algebraic structure of LQFT since one no longer finds commuting subalgebras associated with spacetime regions.  It also
provides additional justification for the locality bounds of \refs{\locbd,\LQGST}, parameterizing breakdown of LQFT.  Section three then discusses other aspects of structure, namely symmetries and evolution, and extra structure, such as the S-matrix, relevant to asymptotically-Minkowski states.  It also discusses a possible connection to the question of unitarization of black hole evolution, and comments both on the challenges to a fundamental role for entanglement, and in particular to the idea that spacetime emerges from entanglement.  Section four discusses how algebraic structure could be a pertinent focal point for the AdS/CFT correspondence, if this correspondence can ultimately be shown to reproduce the details of quantum gravity in anti de Sitter space.  Section five closes with conclusions.

\newsec{Synopsis: algebraic structure of local quantum field theory}

To gain understanding of how one might formulate a 
 a quantum theory in which spacetime emerges in an approximation, it is first useful to review how LQFT, in which an underlying spacetime structure is instead assumed as a foundational ingredient, is described in an intrinsically quantum-mechanical fashion.  The reason is two-fold.  First, such a description of LQFT serves as a possible model for the kind of structure we require in a more general theory.  Second, one expects that a more fundamental quantum theory of gravity should match onto the LQFT description in a {\it correspondence limit}, where one considers, {\it e.g.}, low energies and weak gravitational fields.  As we will describe in this section, an appropriate quantum description of LQFT is  provided by Algebraic Quantum Field Theory, also known as Local Quantum Physics.\foot{For a basic reference, see \refs{\Haag}.}  Local Quantum Physics nicely describes both the quantum-mechanical properties of LQFT, and their interrelation with the locality structure of spacetime.

A well-known ``folk theorem" holds that the assumptions of locality, Poincar\'e invariance, and quantum mechanics imply the basic structure of LQFT.  
Specifically, the Local Quantum Physics approach begins with a notion of an underlying classical spacetime manifold, commonly Minkowski space.  It moreover assumes the standard quantum structures of a Hilbert space $\calh$, and observables which are self-adjoint operators acting on $\calh$.  The Poincar\'e group describes the symmetries of Minkowski space, and as discussed by Wigner, these symmetries (and any internal symmetries) are implemented by a ``ray representation" of the relevant groups.  Additional assumptions are commonly made about the spectrum of the momentum operators $P^\mu$, specifically that this spectrum is restricted to the closed forward light cone, and commonly, that there exists a unique ground state.

Locality itself is encoded in the algebra of observables, whose basic building blocks are the field operators.  Specifically, the classical spacetime geometry can be divided up into subregions, and to each such subregion, one can associate an operator subalgebra, {\it e.g.} thought of as fields smeared with functions with support only in that subregion.  Then, operators associated to subregions that are spacelike separated commute (or, for fermionic operators, anticommute).  

This net of subalgebras, with inclusions of subalgebras corresponding to smaller subregions, therefore mirrors the basic topology of the underlying manifold and the locality property of the  LQFT on it.  The underlying manifold can be reconstructed from this net of subalgebras.  

For finite or locally finite quantum systems,  one often goes a step farther, and describes a factorization of the Hilbert space $\calh$ into corresponding tensor factors on which the different subalgebras act.  Thus, these tensor factors also can be thought of as corresponding to subregions.  In models for the structure of quantum gravity, such constructions have also been used, for example explicitly in discussions of black hole evolution, where the black hole and its surroundings might be thought of as separate tensor factors\refs{\Mathurrev\SGmodels\BHQIUE\GiShI-\Susstrans}.  

However, it turns out that such a tensor factor structure is problematic even in LQFT; as we will discuss, quantum gravity appears to present even greater challenges.

In short, there are two problems with describing factorizations of $\calh$ corresponding to subregions in LQFT:  the first is the type-III property, and the second is the existence of long-range gauge fields.  One might hope to avoid these problems by introducing a regulator, such as the lattice, but we will return to problematic aspects of that momentarily.

The type-III property refers to classification of von Neumann algebras, where the algebras associated with LQFT always lead to type III subfactors.\foot{For completeness: if the space of bounded operators on $\calh$ can be written as a von Neumann algebra $\cala$ together with its commutant, $\calb(\calh)=\cala\vee \cala'$, then $\cala$ is called a {\it factor.}  A type III factor $\cala$ is a factor such that for every projection $E\in \cala$, there exists a $W\in \cala$ that maps $E\calh$ isometrically onto $\calh$, in other words $W^*W=1$, $WW^*=E$.  For more details see \refs{\YngIII}.} Type III subalgebras do {\it not} yield tensor product factorizations of the Hilbert space.   The basic problem can be illustrated by discussing two-dimensional Minkowski space in a Rindler description.  If this has metric
\eqn\tdmink{ds^2=-dt^2+dx^2\ ,}
the two regions $x>|t|$ and $x<-|t|$ are complementary Rindler wedges, with which we could associate subalgebras of the field algebras of a LQFT on this space. The subalgebras correspond to operators acting on the left or right Rindler wedge.  However, the Hilbert space does not correspondingly factorize.  Formally, one can introduce states $|a\rangle$ and $|{\hat a}\rangle$ that describe Rindler excitations of the Rindler vacua for the right or left Rindler wedge, respectively, and formally one can write the Minkowski vacuum in the form
\eqn\minkvac{|0\rangle_M = \prod_\omega S(\omega)  |{\hat 0}\rangle |0\rangle\ }
where $S(\omega)$ is a unitary operator generating squeezed states\CavesZZ\ for modes of frequency $\omega$:
\eqn\sqedef{S(\omega) = \exp\left\{ {z(\omega)} \left( {\hat b}_\omega^\dagger b_\omega^\dagger -  {\hat b}_\omega b_\omega\right)\right\}\ ,}
with
\eqn\zdef{\tanh z(\omega) = e^{-\beta\pi \omega/2}\ }
and $\beta=2\pi$.
Indeed, an even simpler model for such a state is given in the ``qubit approximation\refs{\Mathurrev,\SGmodels},"\foot{This may be obtained from \minkvac\ by assuming that we are treating fermions, and taking the infinite temperature limit, $\beta\rightarrow0$.} where one thinks of each Rindler mode as either occupied, $|1\rangle$, or unoccupied, $|0\rangle$, and a toy model for the Minkowski vacuum is
\eqn\minktoy{|0\rangle_M = \prod_{i=1}^\infty \left( {|{\hat 0}\rangle|0\rangle +  |{\hat 1}\rangle|1\rangle\over \sqrt 2}\right)_i\ ,}
where the index $i$ labels the different modes.  The problem is the infinite entanglement manifest in the expressions \minkvac, \minktoy:  unentangled states of the form $|{\hat \psi}\rangle |\psi\rangle$ are {\it not} in the Hilbert space; they have infinite excitation, and, for example, if one introduces a typical field theory Hamiltonian, they have infinite energy due to the infinite amount of ``broken" entanglement.  This means there is no well-defined factorization $\calh = \calh_L\otimes \calh_R$.  It has been shown that this type-III behavior is the generic behavior of LQFT.\foot{See \BuchholzGR, and other references in \refs{\YngIII}.  For related comments see \RequardtKJA.}

The second problem is associated with long-range gauge fields.  For example, in a gauge theory such as quantum electrodynamics, it is impossible to create a charged particle without creating its long-range field.  So, if we for example consider dividing a region of space into two subregions, generic excitations of charged particles in one of the two subregions will have an effect on the other subregion:  field lines cross the boundary between the subregions.  One approach to this problem is to introduce additional, charged, degrees of freedom on the boundary between the subregions\refs{\Donnelly}, which {\it e.g.} can effectively ``screen" the charges.  However, that changes the basic content of the theory.  One may impose additional conditions on these surface charges, roughly that those from the two sides cancel, but that reintroduces a non-trivial entanglement structure between the two subregions.  Thus, one has no way to write the original Hilbert space $\calh$ as a tensor product $\calh=\calh_1\otimes \calh_2$.  

Of course, a regulator that  makes the number of degrees of freedom finite seemingly has the potential to overcome the type-III problem, and also simplifies the description of the decomposition of a Hilbert space for a gauge theory by introduction of surface charges.  A specific example is regulating by introducing a lattice.\foot{For a discussion of factorization and entanglement in lattice gauge theory, see \CHR.}  A problem with such a lattice or other regulator is that it generically badly breaks the symmetry, specifically Lorentz invariance.  This indicates that, \eg\ in the Rindler case, one has truncated an infinite number of the states of the theory that otherwise would be legitimate states (for example, highly-boosted particles, whose presence is implied by Lorentz invariance).  The result of such a procedure shows up in divergent contributions to quantities like entanglement entropy, as the regulator is removed.  

Breaking the symmetry seems particularly problematic in the context of describing the evolution near a black hole horizon, where the dynamics naturally forces one to consider states with very large relative boosts, like those between infalling matter and outgoing modes that become Hawking radiation.  Indeed, the assumption that such states that are ``disentangled" at a cutoff scale such as the Planck scale can play a role in the dynamics leads to singular horizons\refs{\Trieste,\SGmodels} and is a significant part of the apparent misconceptions that lead to the proposal that physical black holes have such singular, or ``firewall," behavior\refs{\AMPS,\AMPSS}.

In short, in LQFT there is apparently no natural way to decompose the Hilbert space into tensor factors, without drastically modifying the short-distance structure and symmetry of the theory, and this problem becomes even more acute in theories with gauge fields.
This is particularly pertinent to discussions of entanglement, which have been popular in the recent literature.  As we will describe further below, entanglement is most clearly defined relative to a subsystem structure given by such a particular tensor factorization of the Hilbert space, and the absence of such a natural factorization structure appears to be a challenge to a fundamental role for such entanglement.

The basic structure in which a physical system has different subsystems is particularly important in physics, and appears to be manifest in the physical world; degrees of freedom here appear independent of those on Proxima Centauri.  Moreover, such statistical independence of subsystems is central to discussions of measurement, where the world has for example subsystems  corresponding to the physical system being measured, and the observer.  But, this locality and independence of subsystems can fortunately be characterized in terms of the {\it algebraic} structure we have noted,\foot{Often one wants to restrict to the algebras of bounded operators.}  where operators corresponding to different subsystems commute, and apparently doesn't require tensor factorization of the Hilbert space.  This is a rationale for focusing on this algebraic structure: the subalgebras describe the centrally important subsystem structure, and the net captures the topology of the spacetime.

LQFT, via the Local Quantum Physics description, thus provides an example of a quantum theory with subsystem structure which matches well our observations of the locality of physics and also our need for independence of physical systems in order to describe interactions in simple terms, and for a basic description of measurement.  Moreover, as noted, this algebraic structure can be used to reconstruct the underlying spacetime manifold.  This thus furnishes us with an example of the type of quantum structure needed to do physics; when we turn to gravity we do not however expect the quantum structure to be based on an underlying classical geometry, so this structure needs to  be generalized.  

\newsec{Algebraic approach to Quantum Gravity\foot{A natural temptation is to refer to the approach advocated in this paper as ``Algebraic Quantum Gravity," but this name is already taken by the different approach of \refs{\Thie}.}}

In confronting the puzzles of quantum gravity, we assume that the fundamental framework for physics is quantum mechanics, suitably generalized.  In particular, since a priori introduction of space and time concepts is not necessarily warranted, one needs a general set of quantum postulates, such as those of universal quantum mechanics\refs{\UQM}, which do not rely on assuming an underlying space, time, or notion of ``histories\refs{\Hart}."  Specifically, we will assume the following postulates to begin to lay a quantum foundation for physics:

\subsubsec{Q1}  Physical configurations will be described by rays $|\psi\rangle$ of a Hilbert space\foot{Usually expected to be separable, \eg\ to permit thermal ensembles.} $\calh$.  

\subsubsec{Q2}  The space $\calh$ has an inner product, which is (anti)linear under addition and multiplication; the product of two states is denoted $\langle \psi'|\psi\rangle$.

A key question is what additional structure needs to be provided on $\calh$, in order to describe physics. A critical ingredient for physics, generalizing locality, is  that of statistical independence of different subsystems of a physical system, which is important for describing interaction, measurement, {\it etc.}
In finite or locally finite quantum systems of interest to physics, a tensor product decomposition of $\calh$ exhibits this statistical independence and 
is a usual prerequisite for describing interaction and  measurement, as well as entanglement, {\it etc.}  However, as discussed above, in LQFT, we do not expect such a product decomposition of $\calh$ to exist, due to the type-III property and the existence of gauge fields, unless we introduce additional structure that does violence to the physics.  Fortunately, the net of operator algebras  of the theory comes to the rescue, and describes the necessary structure to describe statistically independent, or local, subsystems.  

In treating gravity, let us assume that the basic long-distance/low-energy properties of the theory are indeed approximately well-described by general relativity, but that our job is to find a modification of the structure at short distances and high energies (thus, also long distances) so that we have a fully quantum-mechanical theory.  This means that we can take as a key guide the {\it correspondence} of such a more basic theory with general relativity in the  long-distance/low-energy limit.  In that limit, gravity is of course described as a field theory with a gauge symmetry, that of diffeomorphism invariance.  This suggests that, in introducing a structure such as that of statistically-independent subsystems, we encounter similar challenges as in LQFT, either due to the type-III behavior, or associated with long-range gauge fields.  

Gravity introduces further puzzles.  First, we don't expect there to be a notion of an underlying classical spacetime on which to ``anchor" the net of operator algebras.  A related problem regards locality.  In a gauge theory such as QED, we can formulate local observables which have an effect only on a localized subsystem by considering action of the gauge invariant field strength $F_{\mu\nu}$, of localized Wilson lines, or even of operators that create a particle-antiparticle pair together with a localized field configuration:  these operators, with no net charge, are not required to modify the asymptotic field.

But  a particle is inseparable from its gravitational field, and screening of gravity is problematic.  Any localized excitation is expected to have some non-zero energy, and creation of such an excitation therefore modifies the long-range gravitational field.   We can create a zero-energy operator by integrating a field operator, or product of field operators, over all spacetime, and might hope this avoids gravitational coupling (though this is not necessarily true, if gauge invariance is also required -- see \refs{\GMH}) -- but then one is no longer considering a localized operator.

Our general approach to a quantum description  indicates we should focus on the algebra of observables, and look for interesting refinement of the structure of this algebra that characterizes the physics.  At long distances/low energies we expect these observables to approximate those of LQFT, but gravity has new properties that tell us that the full algebraic structure is expected to be  different from that of LQFT.  First, we do not expect it to incorporate an underlying spacetime structure; instead we might expect that, with quantum mechanics and Hilbert space now in a fundamental role, locality, spacetime, and the equivalence principle emerge from this more basic structure, in the correspondence limit.  Second, if we do assume that general relativity gets the long-distance gauge structure of the theory approximately ``right," then we do not expect to have the same kind of notion of local commuting operators -- even in the correspondence limit, the algebraic structure of the theory is different from that of LQFT.  If, in turn, the algebra of operators is used to {\it define} the underlying analog of spacetime, we expect a very different fundamental structure to ultimately emerge.

One focus of this paper is such preliminary exploration of the possible role of this algebraic structure, and to investigate the novel features, in comparison with LQFT, of this algebraic structure that can be inferred from the long-distance/low-energy behavior of gravity.  If the approach of trying to find a fundamentally quantum theory that matches these long-distance/low-energy properties is a viable one, these should serve as critical clues about, and moreover, constraints on, such a fundamental theory.

\subsec{Algebraic structure and (non)locality}

We begin with another postulate for the quantum theory of gravity:

\subsubsec{A} There is an algebra $\cala$ of ``observables;"  these are self-adjoint operators that act on $\calh$.  

Some comments are in order.  First, we assume that the operators in this algebra are gauge invariant, that is they are left unchanged by  local gauge symmetry or whatever supplants it.  Second, such operators can play different roles in a theory.  They can be used to characterize states, for example by finding eigenvectors and eigenvalues of operators, which may correspond to definite attributes of a state.  They can also be used to {\it change} states, for example by mapping from one state of $\calh$ to another, or if there is unitary evolution generated by a hamiltonian, in which such operators can serve as elementary ``interaction" terms.  

As we have reviewed, in LQFT/Local Quantum Physics, the analogous algebra is further refined by a subalgebra structure, and locality (or more generally statistical independence) is encoded in the commutativity of subalgebras associated with spacelike separated regions.  Therefore, it is clearly of great interest to identify properties of the algebra $\cala$ relevant to a theory of quantum gravity.  Of course, a simple subalgebra structure encoding locality in gravity is more problematic, as we have described above and will see further below.  But nonetheless, we would like to know to what extent there is analogous structure, and also how the subalgebra structure of LQFT emerges in an approximation to that structure in the correspondence limit.

A lot is clearly unknown about the algebraic structure of quantum gravity, but if we do assume that it has a long-distance/low-energy limit that matches onto general relativity, or a perturbatively quantized version of it in that limit, that is an important guide to the more basic structure.  For example, if we discover that certain operators do not commute even in the ``field theory" limit, we expect that lack of commutativity to mirror the more fundamental behavior of the theory.

As we have described, $\cala$ apparently contains no operators that are local in the LQFT sense.  Consider excitations built on  a quantum state corresponding to a large, semiclassical spacetime, such as Minkowski space.  As noted, any localized excitation is expected to have some energy, and thus to alter the long-distance gravitational field.  For that reason, the full operator creating that excitation, together with its inevitable gravitational field, is not local.

This can be made rather precise in the weak-field limit of gravity.\foot{For a more detailed version of the following discussion, see \refs{\DoGi}.}  In fact, even quantum electrodynamics (QED) exhibits some similar behavior -- though with key differences.  Let us begin with QED.  If we have a scalar field $\phi$ with charge $q$, creation of a quantum of this field will be accompanied by creation of its long-range electromagnetic field.  There are many such possible field configurations that can be created, satisfying the Gauss' law constraint.  One such configuration is created by a simple Wilson line, say running in the $z$-direction:
\eqn\wilsEM{\Phi(x)=\phi(x)\exp\left\{i q \int_{z_x}^\infty dz A_z\right\}\ .}
This is gauge invariant under
\eqn\gaugexm{\phi(x)\rightarrow e^{iq\Lambda(x)}\phi(x)\quad ;\quad A_\mu\rightarrow A_\mu + \partial_\mu \Lambda\ .}
This operator however fails to commute with the electric field $E_z$ all along the line, and also two such operators do not always commute at spacelike separations.  Another more symmetric dressing is obtained\DoGi\ by averaging over directions of the Wilson line, to produce the Coulomb field.\foot{More precisely, this produces the dressing of Dirac\refs{\Dirac}; a more general dressing is relevant for non-static charged particles.}  However, this makes the situation worse:  while for a collection of operators of the form \wilsEM\ we can in some cases arrange their electric strings not to intersect, so that the operators commute, averaging over all directions means that there are always contributions to  commutators.

Of course, in QED, or more general non-abelian gauge theories, we can, as noted above, still formulate observables that are strictly localized to a region.  An example is of the form
\eqn\twochage{\cald(x,y) = \phi(x)\exp\left\{i q \int_{z_x}^{z_y} dz A_z\right\}\phi^*(y)\ ,}
which only creates a non-trivial field in a neighborhood of the string connecting $x$ and $y$. Thus, $\cald(x,y)$ commutes with observables spacelike-separated from this neighborhood. Of course note that once created, the ``electric string" along $z$ quickly decays -- the field relaxes to a different electric field\refs{\Shab}, which for static charges would be the combined Coulomb fields (dipole field), and extends to infinity.  Nonetheless, this existence of localized (and even local) gauge invariant operators in QED is one of the key differences with gravity.

Such constructions can be generalized to gravity\refs{\Heem,\KLdecode,\AlmheiriLWA,\DoGi}.  These are most easily explored in the linearized approximation.  Consider, specifically, linearization about Minkowski space,
\eqn\minkpert{ds^2 = (\eta_{\mu\nu} + h_{\mu\nu}) dx^\mu dx^\nu\ .}
The linearized gauge symmetry is 
\eqn\lingauge{x^\mu\rightarrow x^\mu+\xi^\mu\quad ,\quad h_{\mu\nu}\rightarrow h_{\mu\nu} - \partial_\mu \xi_\nu - \partial_\nu \xi_\mu\ .}
One can construct an analog of the Wilson-line dressing \wilsEM\ by defining\DoGi\ (for a related construction, see \refs{\Heem,\KLdecode})
\eqn\wilsgrav{\Phi(x) = \phi(x^\mu+V^\mu(x))}
where
\eqn\Vdef{V_z(x) = -\hf\int_{z_x}^\infty dz h_{zz}\quad , \quad V_{\hat \mu}(x) = -\int_{z_x}^\infty dz \left( h_{z\hat \mu}+\hf \partial_{\hat \mu} \int_z^\infty dz' h_{zz}(x^{\hat \nu}, z') \right)\ ;}
here $\hat \mu$ denotes indices excluding $z$.  This dressing can be shown to be gauge invariant under \lingauge, and creates a ``gravitational string" extending to infinity along the $z$ direction.

The operators \wilsgrav\ have analogous properties to those for QED, \wilsEM\DoGi.  They in general do not commute with the canonical conjugates to $h_{\mu z}$, and thus with curvatures, and also they do not in general commute with each other.  Like in QED, they are expected to create a state that decays to a ``more natural" field configuration.  However, the resulting Coulomb-like operator will generally have nonzero commutators with another such operator.

In fact, one can average such Wilson line operators to find an analog of the Dirac dressing of electromagnetism; this symmetrical dressing creates a linearized version of the Schwarzschild solution, at spacelike separations from the point $x$ at which the particle is created, and a boosted version is expected to create a linearized version of the Aichelberg-Sexl solution.  Properties of these operators and their extension to those for moving particles will be further described in \DoGi.

Since all localized excitations are expected to carry energy, and gravity can't be screened, there is no clear analog of the uncharged operator \twochage,\foot{Though, one may create a particle at one point and annihilate one at another.} and moreover, even scalars, such as the Ricci scalar $R(x)$, are not invariant under \lingauge.  This absence of any local or even localized operators appears to be a very important difference between the algebra $\cala$ of gauge-invariant operators for gravity, as compared to the algebras for LQFT.

Since locality seems quite important, at least in low-energy physics, and since we have seen that an approach to describing locality is through the existence of commuting subalgebras, let us further explore the extent to which we might find such commuting subalgebras.  
	
For example, if we have two operators \wilsgrav\ at the same $x^{\hat \mu}$ but at different $z$, but if we run their Wilson lines in {\it	 opposite} directions along the $z$ axis, so that they don't meet, then these commute in the linearized theory.  Moreover, in the linearized theory we can act with multiple copies of these operators, maintaining commutativity.  Indeed, we could consider such lines terminating in spacelike separated regions, still with commutativity in the linearized approximation, and thus apparently build up commuting subalgebras.

However, we do not expect this construction to hold in the non-linear theory, where the gravitational strings, even if ``thickened," have energy, and where a large number of them has a large amount of energy.  Ultimately we expect commutativity to break down.  Moreover, even in the linear theory, as we have noted, the gravitational field configurations created by operators \wilsgrav\ are unstable; nonlinearities are also expected to contribute to this.  The gravitational field lines will spread out, and in the non-linear theory, cannot get too close.

There is a natural conjecture about when such gravitational strings become too dense to commute in the full theory.  Specifically, if we consider a region of radius $R$, into which such gravitational lines enter from different angular directions, we expect failure of commutativity when the energy $E$ in the region is such that the radius is less than the Schwarzschild radius, $R\leq R(E)$,
with
\eqn\schwR{R(E) = (GE)^{1/(D-3)} }
in $D$ spacetime dimensions.  This would offer further explanation of the ``multi-particle locality bound\refs{\locbd,\LQGST}," which states that an N-particle state fails to be well-described by LQFT when the greatest separation between the particles is less than the Schwarzschild radius corresponding to the total center-of-mass energy of the particles.

Indeed, this is supported by a related argument.  If we consider the electromagnetic dressing of two Coulomb-dressed charged particles separated by a radius $r$, the nonzero commutator of the two operators is of size $q^2/r$.  We expect the same behavior in gravity, with commutator of size $GmM/r^{D-3}$ (again working in general dimension) for two massive particles with masses $m$ and $M$, and this is found from the gravitational dressing\DoGi.  This is expected to be an important modification to the dynamics of the mass $m$ particle when $GM/r^{D-3}=1$.  We also expect two highly-boosted particles, that source Aichelburg-Sexl dressings, and collide from opposite directions, to initially be described by commuting operators.  However, once they reach the region where their separation is of size $R(E)$, with $E$ now the center of mass energy, we expect ultimate failure of commutativity.

So, it appears that even a pair of operators of the form \wilsgrav, with gravitational strings run in opposite directions, do not provide true subalgebras in the nonlinear theory.  If we act multiple times on the vacuum with a single operator \wilsgrav, we expect the commutant to shrink with increasing powers of the single operator.\foot{It also appears that such commutativity is state-dependent, in the sense that two operators may appear to have larger commutators when the commutators are evaluated in certain states.  That is, commutators of two operators may only be small when evaluated on certain states, {\it e.g.} ``near" to the vacuum.}

Similar behavior is expected to appear for other gauge-invariant operators, such as the relational observables considered in \GMH.  There, for similar reasons, we don't expect to find precise subalgebras corresponding to some notion of ``regions" of spacetime.

So already, at the linear level, or in its extension into the nonlinear theory, we appear to be discovering something very important about how the algebraic structure of quantum gravity is quite different from that of LQFT.  Specifically, we don't appear to necessarily find an analogous subalgebra structure to that of LQFT, corresponding to localized excitations in different regions of spacetime.  Such a structure appears to only obviously arise approximately in limits, where gravity is behaving weakly.   We can attempt to characterize this {\it approximate} algebraic structure, by introducing a correspondence postulate:

\subsubsec{C}  The algebra $\cala$ contains operators $\calo_I$ that, in the weak-field limit of low-energies/long distances, approximately reproduce correlation functions computed in LQFT, \eg
\eqn\corresp{\langle 0| \prod_I \calo_I |0\rangle \approx \langle 0| \prod_I O_I |0\rangle_{LQFT}\ ,}
where the $O_I$ are operators of LQFT.

An important question is to better characterize the full algebraic structure, extending it consistently into the full quantum theory, and to understand what further refinements of this structure, such as subalgebras, exist.  Another important goal is 
 to understand  how the local algebras of LQFT emerge from it in such a correspondence limit.

\subsec{Further structural aspects}

While the algebraic structure, or approximate algebraic structure, of a theory of quantum gravity is, as we have argued, very important in better understanding its nature, and requires further study, we turn to a brief discussion of other structure useful in describing such a theory.

Often in physics symmetry plays an important role.  The symmetries that are present will depend on the vacuum on which we work, \eg\ one corresponding to Minkowski space, anti de Sitter space, de Sitter space, or one with less symmetry.  When such a symmetry group is present, we postulate

\subsubsec{Sy}  As described by Wigner, a symmetry is implemented by a representation up to phase of the symmetry group on $\calh$.  

An example of the possible role of such symmetry will be discussed momentarily.

What other structure is necessary for physics?  Of course, typically, as we have emphasized, one begins with a subsystem structure, or in LQFT, a corresponding network of subalgebras.  Then one can describe evolution, \eg\ through interaction of subsystems.  In some contexts this evolution may be described by considering 
 unitary transformations
\eqn\unitaryev{U=e^{-i\lambda H}}
for certain self-adjoint operators $H\in\cala$ and parameter $\lambda$.  For example, if subalgebras can be identified, a basic interaction term may take the form $H=O_1 O_2$ where $O_1$ and $O_2$ lie in commuting subalgebras, corresponding to different degrees of freedom.  Such evolution can be viewed as producing correlation between these degrees of freedom, and is a prototype for interaction, and measurement.  But, this is far too general and we need a clearer understanding of how any such evolution is precisely described in the gravitational context.  Note in particular that, of course, since the weak-gravity limit of evolution is expected to match that governed by the Wheeler-de Witt equation, any evolution may only be apparent evolution, as described through relational observables contained in $\cala$.  So, once again, the key questions seem to return to the structure of the algebra $\cala$.

The preceding comments seem particularly pertinent in cases of closed cosmologies, such as de Sitter space, where identification of the correct relational observables seems an important way to make progress in describing evolution\refs{\GMH,\GiMa,\GiSl}.  Here we don't have the ``platform" of an asymptotic spatial infinity with respect to which we can localize, and so statements must be made with reference to other features of the state, \eg\ ``the spectrum of scalar fluctuations looks like X at the reheating time determined by the value of the inflaton field."

On the other hand, we can also consider states which are expected to have a notion of spatial infinity, such as those which correspond to asymptotically anti de Sitter or Minkowski boundary conditions.  Here we make some comments on the latter, which also have  extensions to the former.  (Further comments on the AdS/CFT correspondence will also be made below.)

\subsubsec{Asymptotically Minkowski states}

If quantum gravity has a sector of states that can be thought of as corresponding to geometries with asymptotically Minkowski boundary conditions, we expect significant extra structure in this sector that may be useful in further elucidating fundamental properties of the theory.  Specifically, according to postulate Sy, we expect these states to transform under the Poincar\'e group.  If so, states may be assigned definite momenta and energies.  (We may also seek to describe these states as evolving with respect to time at infinity.)

There is a further subtlety here.  As described above, {\it any} excitation of the ground state corresponding to the Minkowski vacuum is expected to modify the effective gravitational field at infinity. Far from a localized finite energy state, this field will be small, \eg\ of typical size $GE/r^{D-3}$ at a distance $r$ (defined in the asymptotic metric).  Note, however, that some care is needed in orders of limits.  A large boost may be applied to such a state, and increases its energy and thus asymptotic field.  So, there is not an action of the Poincar\'e group on asymptotically Minkowski states,\foot{One may define asymptotically Minkowski states as those states which are well-described by LQFT near infinity.} except in the sense that for any finite boost, one has to work at sufficiently large radius so the field remains weak.  The limit of infinite boost produces a strong field extending to infinity, and in this sense an infrared cutoff on the theory also serves as an ultraviolet cutoff.

Nonetheless, we might expect that the states of the theory do fall into representations of the Poincar\'e group.  We might also make the following assumption about the spectrum:

\subsubsec{Sp} The spectrum of the momenta $P^\mu$ in the physical Hilbert space $\calh$ is confined to the closed forward light cone, $p^0\geq |{\vec p}|$.  Typically, we also assume the existence of a unique ground state, $|0\rangle$.

In particular, matching onto the known long-distance structure of quantum physics, we expect that there is a subspace of single-particle states that furnishes a representation of the Poincar\'e group.  Moreover, we expect that the ``multi-particle" states of the theory include states where we have two or more such single particles that are widely separated, so their interactions can be neglected.  This  identification of asymptotic  states in terms of particle states is the traditional assumption needed for formulating an S-matrix description of the theory.\foot{Note that for $D=4$, one must still exercise care in treating massless photons and gravitons, and a true S-matrix does not exist.  To avoid this, one may work at higher D, properly treat the gravitational and electromagnetic dressings described above, or formulate appropriate inclusive scattering probabilities.}

These asymptotic states are expected to be characterized by their momenta and other quantum numbers, and the S-matrix is expected to have analytic structure associated with appropriate infrared behavior of the theory, approximately matching onto that of local quantum field theory.  In gravity, it is also expected to have other new features capturing the long-range behavior of gravity\refs{\GiSr\GiPo-\Erice}, and differing from standard LQFT.  Systematic study of the analytic behavior of and constraints on the gravitational S-matrix may provide important clues to the more fundamental behavior of quantum gravity.

\subsec{Possible implications of algebraic structure}

The preceding discussion has reviewed how even in LQFT one is limited in defining subsystems, since precise tensor factors can't be defined; we then have found that with gravity, departures from traditional local structures are even more profound, as anticipated in \refs{\locbd,\LQGST}.  This ultimately appears to conflict with a  statement such as the equivalence principle, which assumes that a physical subsystem may be isolated from its gravitational surroundings, and, if small enough, behaves like a system in empty flat space.  The remainder of this section will discuss some other possible implications of the preceding discussion.

\subsubsec{Unitarity of black hole evolution}

An underlying motivation of part of the preceding discussion is to understand how black hole evolution can be unitary; unitarity, here, can \eg\ be defined as that of the S-matrix describing formation and disintegration of a black hole.  For black holes, LQFT apparently utterly fails to capture the correct dynamics, and appears to predict a massive violation of unitarity\refs{\Hawkinc}.  What is apparently needed is evolution that, with respect to the semiclassical description of the black hole geometry, nonlocally transfers information\refs{\BHMR\BHIUN-\NLvC,\SGmodels,\BHQIUE,\AMPS,\GiSh\NVNL\NVNLFT\GiShii-\Gimod}.

Interestingly, we have found that in gravity locality is difficult to formulate to begin with; this certainly provides encouragement that lack of LQFT locality is an important part of the story of black hole evolution.  An important question is to more completely characterize evolution, and see how it approximately matches onto that of LQFT, together with corrections sufficient to unitarize black hole formation and decay.    Refs.~\refs{\BHMR,\AMPS} considered a ``harder" form of such transfer, where the new effects responsible for transferring information dramatically modify the physics in the vicinity of the would-be horizon.  Refs.~\refs{\SGmodels,\BHQIUE,\GiSh\NVNL\NVNLFT\GiShii-\Gimod} instead propose a much ``softer" form of effective transfer, where information is transferred from a black hole interior to its atmosphere, on scales determined by the black hole size, in a fashion that an infalling observer might characterize as ``nonviolent."  In particular, such effects may be parameterized in an effective field theory framework in terms of ``small" deviations from LQFT evolution.  It would be satisfying to more closely connect such deviations to the fundamental obstacles to LQFT locality described in this paper.

\subsubsec{Entanglement, and comments on spacetime from entanglement}

Study of entanglement has been au courant, and it has even been proposed that spacetime itself emerges from entanglement.  It is worthwhile to reassess the role of entanglement, given the discussion of this paper.

Entanglement entropy is usually defined in finite or locally finite quantum systems which possess a subsystem structure defined by a tensor factorization of the Hilbert space.  Indeed, specification of such a subsystem structure is {\it necessary} to sharply define entanglement.  For example\ZanardiZZ, consider a Hilbert space spanned by the states of two qubits, and consider the Bell states, $|\Phi^\pm\rangle = (|00\rangle\pm |11\rangle)/\sqrt 2$ and $|\Psi^\pm\rangle= (|01\rangle\pm |10\rangle)/\sqrt 2$, within it.  If asked whether these states are entangled, many would first react ``yes."  But, we can choose another basis, where these can be rewritten as $|\chi^\lambda\rangle= |\chi\rangle |\lambda\rangle$, with $\chi=\Phi,\Psi$, and $\lambda=+,-$.  With respect to this subsystem decomposition, the Bell states are unentangled.  Entanglement is defined with respect to the subsystem structure given by a tensor factorization. 

Already in LQFT one encounters difficulties in quantifying entanglement by entanglement entropy.\foot{Though, other entanglement measures such as relative entropy of entanglement may be well defined.}  This is directly related to the type-III property that we have described, and the corresponding lack of a factorization structure.  Instead, one must truncate the theory with a cutoff, and one finds an entanglement entropy that diverges when the cutoff is removed. 

The situation seems even more problematic in quantum gravity, where we appear to be even further removed from having a sharp definition of a subsystem structure, since we we haven't even yet identified a precise subalgebra structure.  Thus, it seems that more needs to be learned about the algebras relevant to quantum gravity.  If they have a natural subalgebra structure, and if a useful definition of entanglement can be defined relying only on this structure, and doesn't require the Hilbert space to factorize (or if somehow, in gravity, it does), then entanglement entropy might be precisely defined and fruitfully explored.

The proposal that entanglement builds up spacetime\refs{\vanR,\MaSu} is even more puzzling.  It is true that in LQFT, entanglement is implied by the underlying spacetime structure -- the vacuum and finite energy excitations of it are infinitely entangled, as we have described.  But this is far from saying that entanglement somehow suffices to define spacetime.  Indeed, we have just said that defining entanglement requires a subsystem structure, and in LQFT this subsystem structure (to the extent to which it is defined, with a cutoff) is closely associated with the topology (structure of subregions) of the preexisting underlying manifold.  In short, it would appear that subsystem structure, whether defined by tensor factorizations or subalgebras, comes first, and to the extent to which entanglement is precisely defined, it is with respect to this structure.

Indeed, the notion of ``building up" spacetime from entanglement is puzzling from another perspective.  Another place that the spacetime structure appears in LQFT is in governing the structure of the hamiltonian.  If a subsystem structure is defined by subalgebras, then interactions between ``nearby" subsystems are encoded in interaction terms $H_i = O_1O_2$, with $O_1$ and $O_2$ elements of different subalgebras.  Such interactions can alter entanglement between different subsystems, but it is not clear how they would arise from the entanglement, if that was what was responsible for defining the spacetime structure!  Again, spacetime structure appears more primitive than entanglement, and is correspondingly encoded in the hamiltonian that can govern the nature and evolution of the entanglement.  Put differently:  one may propose that an ordinary EPR pair is connected by a spacetime bridge.  But, there is nothing in the hamiltonian of a widely-separated pair that encodes interactions between the partners of the pair.  In this sense, they are neither proximal nor connected.

\newsec{Algebraic aspects and puzzles of AdS/CFT}

It has been proposed that AdS/CFT provides a {\it definition} of quantum gravity with AdS boundary conditions, though for a critical viewpoint see \refs{\Isit,\GaGi}.  Let us, however, momentarily ignore the difficulties found in trying to reconstruct the fine-grained structure of the bulk theory, and ask the question, if AdS/CFT does define a theory of quantum gravity, {\it how} would it do so?

Since the boundary theory is a LQFT, it has the kind of subalgebra structure, associated with boundary subregions, that we have described for LQFTs.  However, it is not clear whether the LQFT subalgebra structure of the boundary CFT is related to the kind of algebraic structure expected for gravity in anti de Sitter space, though there have been conjectures about possible subregion-subregion duality\refs{\BLR\CzechBH\BoussoMH-\HubenyWA}.  As we have noted, when considering gravitationally-dressed operators in bulk physics, we haven't yet identified a natural subalgebra structure, though we expect that in the correspondence limit described above, in a bulk gravity theory we should be able to {\it approximately} recover subalgebras appropriate to describing bulk locality.  An important challenge, which could be a part of the demonstration of how AdS/CFT works, would be to identify this approximate subalgebra structure in the algebra of operators of the boundary theory.  Thus, this identification, particularly for approximate subalgebras corresponding to regions small as compared to the AdS radius scale, could be a key part of establishing the utility of AdS/CFT for describing bulk physics,  if it did indeed describe the detailed structure of bulk physics.

It has also been proposed that two copies of a conformal field theory, entangled in a thermofield double state, leads to an AdS black hole with Einstein-Rosen bridge\refs{\Maldeternal}.  This proposal seems also subject to the preceding comments regarding geometry from entanglement, and thus appears equally puzzling.
		
\newsec{Conclusions}	

This paper has explored a starting point for quantum gravity grounded in quantum mechanics, rather than beginning with spacetime.  This approach is suggested by indications that the usual locality of local quantum field theory is not a fundamental property of gravitational theory, and by difficulties of approaches that begin with spacetime and then try to quantize the metric.
	
If the correct framework for quantum gravity is intrinsically quantum-mechanical, an important question is what mathematical structure is needed beyond that of Hilbert space.  While for finite or locally finite quantum systems important additional structure is supplied by a tensor factor structure for the Hilbert space, such a structure is problematic even in field theory both due to the type-III property (infinite entanglement), together with the presence of long range gauge fields.  Instead, one focuses on the algebraic structure, and a net of subalgebras which correspond to subregions of the spacetime.  Moreover, in gauge theory only certain restricted classes of operators define commuting subalgebras.

These observations prompt exploration of a possible fundamental role for the algebra of observables in quantum gravity, and indicate the importance of understanding further refinements of this algebraic structure.  Given that a particle is inseparable from its gravitational field, and that gravity apparently cannot be screened, one finds an obstacle to even identifying a subalgebra structure associated with regions, a subtlety going beyond that of gauge theory.  This is readily seen if one assumes a principle of correspondence, where the quantum structure of gravity approximately matches onto that of quantized general relativity in the long-distance/low-energy limit.  Even in this limit, gravitationally-dressed operators generally fail to commute even when describing excitations which na\"\i vely are created in spacelike-separated regions.  Thus, this further confirms and quantifies the limitations of local quantum field theory that have  been parameterized by locality bounds\refs{\locbd,\LQGST}.  One does find that commutators can be small, in the long-distance/low-energy limit, so it appears possible to recover the subalgebras of local quantum field theory approximately in the correspondence limit.
	
This discussion appears to have important implications for attempts to find a quantum theory of gravity.  There is not a clear primary role for entanglement, given the difficulty with defining tensor factorizations, and it is difficult to see how spacetime itself could emerge from entanglement.  Moreover, if one takes a quantum information perspective and thinks of particles as roughly corresponding to qubits, and asks the question ``how big is a qubit?" it appears that the answer is that the qubit is arbitrarily large in the sense of having infinitely extended weak field, and moreover its strong-field region grows with the energy of the qubit.  This is a qualitative difference with behavior of more familiar quantum systems.

If quantum gravity can be formulated in such an intrinsically quantum-mechanical framework, it will be very important to further characterize the structure of its algebra of observables, and possible refinement of that structure.  Important guides in this include correspondence with the known long-distance/low-energy behavior of gravity.  Indeed, ultimately one might anticipate that the familiar geometric structure of spacetime emerges from a more basic quantum algebraic structure, defining such a ``quantum emergent geometry."

\bigskip\bigskip\centerline{{\bf Acknowledgments}}\nobreak

I thank J. Hartle, D. Marolf, G. Moore, and J. Yngvason for helpful conversations.  I particularly thank W. Donnelly for valuable discussions, and for collaboration on the upcoming work \DoGi.  This work was initiated at the Aspen Center for Physics Workshop on Emergent Spacetime in String Theory; the Center's hospitality and support of NSF Grant \#1066293 are gratefully acknowledged.
This work  was supported in part by the Department of Energy under Contract DE-SC0011702 and by  Foundational Questions Institute grant number FQXi-RFP3-1330.

\listrefs
\end